\documentclass[prd,onecolumn,reprint,superscriptaddress]{revtex4-2}
\usepackage{feynmp,amsmath,mathtools,amssymb,graphicx,subfigure,bm,mathrsfs,slashed,ulem}
\usepackage{CJK}
\usepackage[utf8]{inputenc} 
\usepackage{color,ulem}
\usepackage[colorlinks=true]{hyperref}

\begin{document}
	\begin{CJK*}{UTF8}{}

		\title{Interband optical conductivity in two-dimensional semi-Dirac bands tilting along the quadratic dispersion}
		
		\author{Xin Chen}
		\thanks{These two authors contributed equally to this work.}
		\affiliation{Department of Physics, Institute of Solid State Physics and Center for Computational Sciences, Sichuan Normal University, Chengdu, Sichuan 610066, China}
		
		\author{Jian-Tong Hou}
		\thanks{These two authors contributed equally to this work.}
		\affiliation{Department of Mathematics,
        The Hong Kong University of Science and Technology, Kowloon, Hong Kong, China}
		\affiliation{Department of Physics, Institute of Solid State Physics and Center for Computational Sciences, Sichuan Normal University, Chengdu, Sichuan 610066, China}

		\author{Long Liang}
		\affiliation{Department of Physics, Institute of Solid State Physics and Center for Computational Sciences, Sichuan Normal University, Chengdu, Sichuan 610066, China}

        \author{Jie Lu}
		\affiliation{College of Physics Science and Technology, Yangzhou University, Yangzhou 225002, China}
		
		\author{Hong Guo}
		\affiliation{Department of Physics, McGill University, Montreal, Quebec H3A 2T8, Canada}
		\affiliation{Department of Physics, Institute of Solid State Physics and Center for Computational Sciences, Sichuan Normal University, Chengdu, Sichuan 610066, China}
		
		\author{Chang-Xu Yan}
		\thanks{Corresponding author: cxyan@mail.bnu.edu.cn}
		\affiliation{School of Physics and Astronomy, Beijing Normal University, Beijing 100875, China}
		\affiliation{Department of Physics, Institute of Solid State Physics and Center for Computational Sciences, Sichuan Normal University, Chengdu, Sichuan 610066, China}
		
		\author{Hao-Ran Chang ({\CJKfamily{gbsn}{张浩然}})}
		\thanks{Corresponding author: hrchang@mail.ustc.edu.cn}
		\affiliation{Department of Physics, Institute of Solid State Physics and Center for Computational Sciences, Sichuan Normal University, Chengdu, Sichuan 610066, China}
		
		\date{\today}

		\begin{abstract}
			
			Two-dimensional (2D) semi-Dirac materials feature a unique anisotropic band structure characterized by quadratic dispersion along one spatial direction and linear dispersion along the other, effectively hybridizing ordinary and Dirac fermions. The anisotropy of energy dispersion can be further modulated through band tilting along either spatial direction of the wave vector. We propose a new definition of tilt parameter to characterize Lifshitz phases in 2D semi-Dirac bands tilting along the quadratically dispersing direction. Using linear response theory, we theoretically investigate the interband optical conductivity of 2D tilted semi-Dirac bands. Our analytical zero-temperature results reveal pronounced distinctions from Dirac and semi-Dirac systems tilting along the linearly dispersing direction. Notably, we find that spectral fixed point emerges in the optical conductivity over a specific range of the tilt parameter, a phenomenon explained by the corresponding behavior of the joint density of states. These findings provide a robust theoretical framework for identifying and characterizing 2D tilted semi-Dirac materials and establish clear spectral fingerprints that distinguish different kinds of 2D semi-Dirac bands and Dirac bands. Our predictions can guide future experimental studies of anisotropic band engineering and tilt-dependent phenomena.
		\end{abstract}

		\maketitle
	\end{CJK*}	
	\newpage
	
	%%%%%%%%%%%%%%%%%%%%%%%%%%%%%%%%
	\section{Introduction\label{Sec:intro}}
	%%%%%%%%%%%%%%%%%%%%%%%%%%%%%%%%

    Fermions are described by parabolic energy-momentum dispersion in conventional semiconductors whereas  by linear dispersion in Dirac and Weyl semimetals  \cite{SCINovoselov2004,RMPCastro2009,RMP2018}. Interestingly, a distinct class of exotic fermions has been predicted in various two-dimensional (2D) materials---including (BEDT-TTF)$_2$I$_3$ salt under pressure \cite{JPSJ2006}, $\mathrm{VO_2/TO_2}$ heterostructures \cite{PRLPardo2009,PRLBanerjee2009}, strained honeycomb lattices \cite{PRBHasegawa2006,PRLZhu2007,NJPWunsch2008,PRBMontambaux2009},  photonic crystals \cite{OEPYWu2014}, and striped boron sheet \cite{PCLZhang2017}---and experimentally observed in black phosphorus \cite{SCIKim2015}. In these materials, the 2D low-energy dispersion is characterized by quadratic dispersion along one spatial direction and linear dispersion along the other, effectively hybridizing ordinary and Dirac fermions, hence termed semi-Dirac bands (SDBs) \cite{PRLPardo2009,PRLBanerjee2009,PRLDietl2008}.  The hybridization of two kinds of distinct fermions dictates intrinsic anisotropic dispersion, leading to unique physical properties \cite{PRLDietl2008,PRXRoy2018,PRBCarbotte2019,PRBZhang2022,PRBOriekhov2022,PRBCarbotte2019B,PRBFekete2025,PRXBasov2025,PRBWang2022,PRBWang2023} that diverge significantly from those of 2D Dirac bands (DBs) exhibiting linear dispersion in both directions.

	Band tilting along specific wave vector directions further introduces anisotropy of energy dispersion and can drive a Lifshitz transition \cite{Lifshitz1960}---a change in Fermi surface topology---with profound implications for material properties \cite{qin2024quantum,PRBTan2022,PRBHou2023,PRBYan2023,PRBRostamzadeh2019,Wild2022optical,PRBYao2021,Xiong2021spin,Tchoumakov2016magnetic,Tchoumakov2017magnetic,Zyuzin2016intrinsic,Ferreiros2017anomalous,Saha2018anomalous,Yu2016predicted,Hou2017double,O2016magnetic,Sonowal2019giant,Das2019linear,Sadhukhan2020novel,Duan2019signature,Tamashevich2022nonlinear,Carbotte2016dirac,Mukherjee2017absorption}. While band tilting in 2D DBs can be uniformly described using a conventional approach, the situation in 2D SDBs is markedly different: tilting along the quadratic dispersion yields qualitatively distinct behavior compared to tilting along the linear dispersion \cite{PCLZhang2017,PRBYan2023,PRBRostamzadeh2019}. Previous studies on tilted bands in both DBs and SDBs have primarily focused on cases where the tilting occurs along the linear dispersion, employing a tilt parameter defined as $t=v_t/v_F$ \cite{PRBTan2021,PRBTan2022,PRBHou2023,JPSNishine2010,PRBMojarro2022,PRBVerma2017,PRBHerrera2019,PRBRostamzadeh2019,PRBMojarro2021,PRBYao2021,PRBZhang2018,NanomaterialsZhu2021,CPBZhu2022,NJPYang2019,JMMMYang2019,PRLTsymbal2023,PRBTan2022,PRBHou2023,PRBYan2023,Wild2022optical,PRBYao2021,Xiong2021spin,Tchoumakov2016magnetic,Tchoumakov2017magnetic,Zyuzin2016intrinsic,Ferreiros2017anomalous,Saha2018anomalous,Yu2016predicted,Hou2017double,O2016magnetic,Sonowal2019giant,Das2019linear,Sadhukhan2020novel,Duan2019signature,Carbotte2016dirac,Mukherjee2017absorption,Tamashevich2022nonlinear,PRBYan2023,PRBRostamzadeh2019,Wild2022optical}. However, this definition is inadequate for describing 2D SDBs tilting along the quadratic dispersion. On the other hand, as a typical transport quantity, optical conductivity can be used to extract the essential information of band structures and optical properties of materials \cite{RMPCastro2009,RMP2018}. Due to the anisotropic dispersion in 2D DBs and SDBs tilting along the linear dispersion, the optical conductivity exhibits highly anisotropy, and the exotic behaviors therein can be used to characterize the Lifshitz transition \cite{PRBTan2021,PRBTan2022,PRBHou2023,PRBYan2023,Wild2022optical,JPSNishine2010,PRBMojarro2022,PRBVerma2017,PRBHerrera2019,PRBRostamzadeh2019,PRBMojarro2021,PRBYao2021,Carbotte2016dirac,Mukherjee2017absorption}. By contrast, the optical response of 2D SDBs tilting along the quadratic dispersion remains largely unexplored. We anticipate pronounced differences in optical conductivity of 2D SDBs tilting along the quadratic dispersion compared to previously studied cases. 
    
    Motivated by this, we propose a definition of tilt parameter for 2D SDBs tilting along the quadratic dispersion. Utilizing it to classify the Lifshitz phases, we theoretically investigate the influence of anisotropic energy dispersion and band tilting on the optical absorption in 2D SDBs. We find that the highly
	anisotropic optical conductivity in 2D SDBs tilting along the quadratic dispersion differs significantly from those in both 2D SDBs and DBs tilting along the linear dispersion. Besides, the interband optical conductivity exhibits a fixed point at $\omega=2\mu$ in 2D SDBs tilting along the quadratic dispersion, similar as those in 2D SDBs and DBs tilting along the linear dispersion, except for different ranges of tilt parameter for the presence of fixed point. In accompany with the findings regarding the fixed point in our previous papers \cite{PRBTan2022,PRBHou2023,PRBYan2023}, we claim that a robust fixed point can appear in the interband optical conductivity of tilted bands for a suitable range of tilt parameter. Our theoretical predictions can be used to fingerprint distinct kinds of 2D tilted SDBs and DBs in optical measurements, and guide future experimental studies of anisotropic band engineering and tilt-dependent phenomena. 
	
	The rest of this work is structured as follows. In Sec. \ref{Sec: Model and Theoretical formalism}, we introduce the model Hamiltonian of 2D SDBs tilting along the quadratic dispersion, propose a definition of tilt parameter, and provide the theoretical formalism for calculating optical conductivity. We present the analytical results of the interband optical conductivity and the joint density of states (JDOS) at zero temperature and the corresponding discussions in Sec. \ref{Sec: Results and Discussions}. In addition, Sec. \ref{Sec: Comparisons and Conclusions} makes a comprehensive comparison and conclusion. Finally, we give two appendixes to show the detailed calculation and analysis.

	%%%%%%%%%%%%%%%%%%%%%%%%%%%%%%%%
	\section{Model and Theoretical formalism	\label{Sec: Model and Theoretical formalism}}
	%%%%%%%%%%%%%%%%%%%%%%%%%%%%%%%%

\begin{figure*}[htbp]
		\includegraphics[width=16cm]{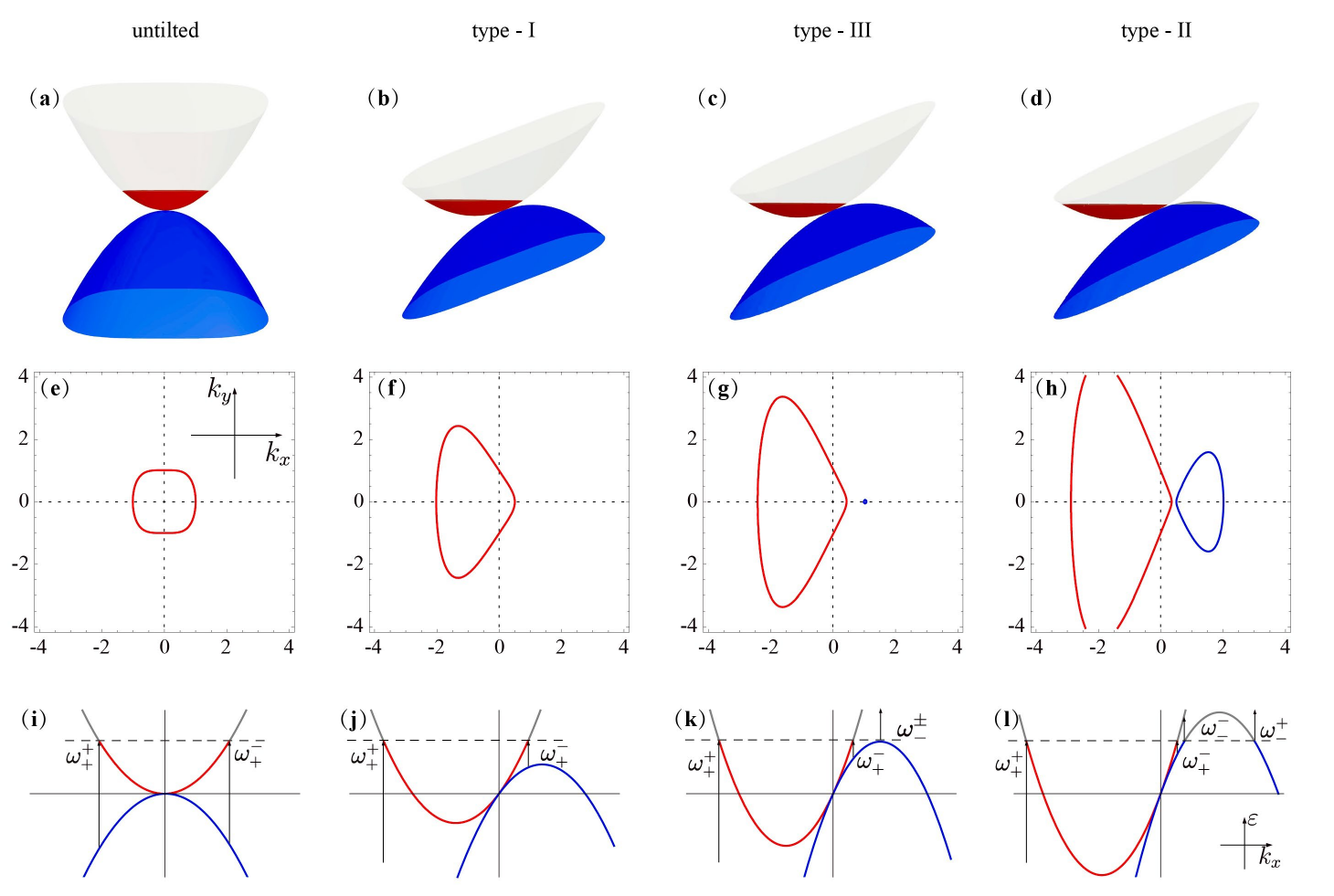}
		\caption{Schematic diagrams for energy bands, Fermi surfaces, and corresponding critical frequencies $\omega_{\lambda}^{\pm}$ of interband optical transitions in 2D SDBs tilting along the quadratic dispersion for $n$-doped case.}
		\label{Figure1}
	\end{figure*}

	We begin with the following effective Hamiltonian for 2D tilted SDBs
	\begin{align}
		\mathcal{H}_\kappa(k_x,k_y)&=\kappa(v_{tx} k_x+v_{ty} k_y)\tau_0
%		\nonumber\\&
		+ a k_x^2 \tau_1+ v_{F} k_y\tau_2,
		\label{Eq3}
	\end{align} 
	where the indices $\kappa = \pm $ denote a couple of valley, $\tau_0$ is $2\times2$ identity matrix, and $\tau_i$ are Pauli matrices acting in pseudospin space. The first two terms describe band tilting along the $x$- and $y$-directions, while the remaining terms govern the intrinsic anisotropy of 2D SDBs \cite{JPSJ2006,PRLPardo2009,PRLBanerjee2009,PRBHasegawa2006,PRLZhu2007,NJPWunsch2008,PRBMontambaux2009,OEPYWu2014,PCLZhang2017}. Hereafter, we set $\hbar=1$ for convenience. The eigenvalues of the Hamiltonian are given by
	\begin{align}
		\varepsilon_\kappa^\lambda(k_x,k_y)=\kappa (v_{tx} k_x+v_{ty} k_y)+\lambda \mathcal{Z}(k_x,k_y),
	\end{align}
	where $\lambda=\pm$ label the electron pocket ($\lambda=+$) and hole pocket ($\lambda=-$), and
	\begin{align}
		\mathcal{Z}(k_x,k_y)=\sqrt{\left(ak_x^2\right)^2+(v_{F} k_y)^2}.
	\end{align}
	The corresponding energy bands and Fermi surfaces are shown schematically in Fig. \ref{Figure1}(a)-(d) and (e)-(h), respectively.
	Obviously, the system exhibits particle-hole symmetry.

	In this work, we propose a tilt parameter
	\begin{align}
		t_{x}&%=\frac{v_{tx}}{v_{xc}}
        =\frac{v_{tx}}{2\sqrt{a|\mu|}},
	\end{align}
	for doped case ($\mu\neq 0$) to characterize the Lifshitz phases in the 2D SDBs tilting along the quadratic dispersion ($v_{tx}\neq 0$ and $v_{ty}= 0$). Within this definition of tilt parameter, the Lifshitz phases are classified as $t_{x}=0$ for untilted phase, $0<t_{x}<1$ for type-I phase, $t_{x}=1$ for type-III phase, and $t_{x}>1$ for type-II phase. This classification is the same as that in 2D DBs and SDBs tilting along the linear dispersion ($v_{tx}= 0$ and $v_{ty}\neq 0$), except for that the definition of tilt parameter therein is $t_{y}=v_{ty}/v_{F}$ \cite{PRBYan2023,PRBHou2023,PRBTan2022}.

	Within the linear response theory, the optical conductivities for the photon frequency $\omega$ and chemical potential $\mu$ is given by
	\begin{align}
		\sigma_{ij}(\omega,\mu,t_{x})
		&=\sum_{\kappa=\pm}\sigma_{ij}^\kappa(\omega,\mu,t_{x}),
	\end{align}
	where 
	\begin{align}
	\sigma_{ij}^{\kappa}(\omega,\mu,t_{x})
	=&\frac{i}{\omega}
	\frac{1}{\beta}\sum_{\Omega_m}\int_{-\infty}^{+\infty}\frac{dk_x}{2\pi}\int_{-\infty}^{+\infty}\frac{dk_y}{2\pi}
	\nonumber\\&
	\mathrm{Tr}\left[\hat{J}_{i}^{\kappa}G_{\kappa}(k_x,k_y,i\Omega_m,\mu)
	\hat{J}_{j}^{\kappa}
	\right.\nonumber\\&\left.\hspace{0.5cm}
	G_{\kappa}(k_x,k_y,i\Omega_m+\omega+i\eta,\mu)\right],
	\end{align}
	with $i,j=x,y$ referring to spatial coordinates, $\eta$ denoting a positive infinitesimal, and $\beta=1/k_B T$ with $k_B$ being the Boltzmann's constant and $T$ being the temperature. The charge current operators read
	\begin{align}
			\hat{J}_x^{\kappa}&=e\frac{\partial \mathcal{H}_\kappa(k_x,k_y)}{\partial k_x}=e \left(\kappa v_{tx}\tau_0 + 2a k_x\tau_1\right),\\
			\hat{J}_y^{\kappa}&=e\frac{\partial \mathcal{H}_\kappa(k_x,k_y)}{\partial k_y}=e v_F\tau_2,
	\end{align}
	and the Matsubara Green's function in momentum space takes the form
	\begin{align}
			G_\kappa(k_x,k_y,i\Omega_m,\mu)
			&=\frac{1}{2}\sum_{\lambda=\pm}
			\frac{\mathcal{P}_{\kappa}^{\lambda}(k_x,k_y)}{i\Omega_m+\mu-\varepsilon_\kappa^\lambda(k_x,k_y)},
	\end{align}
	where $\mu$ is the chemical potential, and
	\begin{align}
			\mathcal{P}_{\kappa}^{\lambda}(k_x,k_y)
			&=\tau_0+\lambda\frac{ak_x^2 \tau_1+v_F k_y\tau_2}{\mathcal{Z}(k_x,k_y)}.
	\end{align}
	After summing over Matsubara frequency $\Omega_m$, we express the optical conductivities for chemical potential $\mu$ at the $\kappa$ valley as
	\begin{align}
		&\sigma_{ij}^{\kappa}(\omega,\mu,t_{x})
			=\frac{i}{\omega}\int_{-\infty}^{+\infty}\frac{dk_x}{2\pi}\int_{-\infty}^{+\infty}\frac{dk_y}{2\pi}
			\sum_{\lambda,\lambda^{\prime}=\pm}
			\nonumber\\&
			\mathcal{F}_{ij}^{\kappa;\lambda\lambda^{\prime}}(k_x,k_y)
		    \frac{f[\varepsilon_{\kappa}^{\lambda}(k_x,k_y),\mu]
		    	-f[\varepsilon_{\kappa}^{\lambda^{\prime}}(k_x,k_y),\mu]} {\omega+\varepsilon_{\kappa}^{\lambda}(k_x,k_y)
		    	-\varepsilon_{\kappa}^{\lambda^{\prime}}(k_x,k_y)+i\eta},\label{DefOCApp}
		\end{align}
		where
		\begin{align}
		  \mathcal{F}_{ij}^{\kappa;\lambda\lambda^{\prime}}(k_x,k_y)
			=\frac{\mathrm{Tr}[\hat{J}_i^{\kappa}\mathcal{P}_{\kappa}^{\lambda}(k_x,k_y)
					\hat{J}_j^{\kappa}\mathcal{P}_{\kappa}^{\lambda^{\prime}}(k_x,k_y)]}{4},\label{FDef}
		\end{align}
		and $f(x,\mu)=1/\left\{1+\exp[\beta(x-\mu)]\right\}$ is the Fermi-Dirac distribution function.

	One can verify that $\sigma_{ij}(\omega,\mu,t_{x})$ respects particle-hole symmetry, i.e.,
	\begin{align}
		\sigma_{ij}(\omega,\mu,t_{x})&=\sigma_{ij}(\omega,-\mu,t_{x})=\sigma_{ij}(\omega,|\mu|,t_{x}).
	\end{align}
	Our analysis is restricted to the real part of the optical conductivity corresponding to optical absorption in the $n$-doped case ($\mu > 0$). After some standard algebra, $\mathrm{Re}~\sigma_{ij}(\omega,\mu,t_{x})$ can be divided into the interband part $\mathrm{Re}~\sigma_{ij}^{\mathrm{(IB)}}(\omega,\mu,t_{x})$ and the intraband part $\mathrm{Re}~\sigma_{ij}^{\mathrm{(D)}}(\omega,\mu,t_{x})$. Because the intraband optical conductivities $\mathrm{Re}~\sigma_{ij}^{\mathrm{(D)}}(\omega,\mu,t_{x})$ contributes only around $\omega= 0$, in the following we focus on the interband optical conductivities given by
		\begin{align} 
		&\mathrm{Re}~\sigma_{ij}^{\mathrm{(IB)}}(\omega,\mu,t_{x}) 
		= \pi \int^{+\infty}_{-\infty}\frac{dk_x}{2\pi} \int^{+\infty}_{-\infty}\frac{dk_y}{2\pi}
		\nonumber\\&\times
        \sum_{\kappa=\pm}\mathcal{F}^{\kappa;-+}_{ij}(k_x,k_y)
		\frac{ f\left[\varepsilon_\kappa^{-}(k_x,k_y),\mu\right] -f\left[\varepsilon_\kappa^{+}(k_x,k_y),\mu\right]}{\omega}
        \nonumber\\&\times
		\delta\left[\omega-2\mathcal{Z}(k_x,k_y)\right],
	\end{align}
	where the explicit forms of the functions $\mathcal{F}_{ij}^{\kappa;-+}(k_x,k_y)$
	are given as
    \begin{align}
		\mathcal{F}_{xy}^{\kappa;-+}(k_x,k_y)
		&=-\frac{2e^2a^2v_F^2k_x^3k_y}{\mathcal{Z}^2(k_x,k_y)},\\
		\mathcal{F}_{xx}^{\kappa;-+}(k_x,k_y)
		&=2e^2a^2k_x^2\left[1-\frac{a^2k_x^4
			-v^2_F k^2_y}{\mathcal{Z}^2(k_x,k_y)}\right],\\
		\mathcal{F}_{yy}^{\kappa;-+}(k_x,k_y)
		&=\frac{e^2v_{F}^2}{2}\left[1+\frac{a^2k_x^4
			-v^2_F k^2_y}{\mathcal{Z}^2(k_x,k_y)}\right].
	\end{align}
	Since $\mathcal{F}_{xy}^{\kappa;-+}(k_x,k_y)$ is an odd function of $k_y$, the interband transverse optical conductivities vanish, namely, $\mathrm{Re}~\sigma_{xy}(\omega,\mu,t_{x}) = \mathrm{Re}~\sigma_{yx}(\omega,\mu,t_{x}) = 0$.
	We now turn to the real part of the interband longitudinal optical conductivities (LOCs) for the $n$-doped case ($\mu > 0$). For sake of analytical calculation, we assume zero temperature ($T=0~\mathrm{K}$), at which the Fermi-Dirac distribution function reduces to the Heaviside step function, $f(x,\mu)=\Theta(\mu-x)$.

		%%%%%%%%%%%%%%%%%%%%%%%%%%%%%%%%
	\section{Results and Discussions	\label{Sec: Results and Discussions}}
	%%%%%%%%%%%%%%%%%%%%%%%%%%%%%%%%
	
	The interband LOCs, $\mathrm{Re}~\sigma_{jj}^{\mathrm{(IB)}}(\omega,\mu,t_{x})$, can be expressed in a unified form,
	\begin{align}
		\mathrm{Re}~\sigma_{jj}^{\mathrm{(IB)}}(\omega,\mu,t_{x})&= S_{jj}^{(\mathrm{IB})}(\omega) \Gamma_{jj}^{(\mathrm{IB})}(\omega,\mu,t_{x}).
		\label{LOCsp1}
	\end{align}
	The
	dimensional magnitudes $S_{jj}^{(\mathrm{IB})}(\omega)$ reflect the
	scaling of dimensional dependence with respect to frequency $\omega$ in the interband LOCs, and are defined as
	the undoped ($\mu=0$) interband LOCs for 2D untilted SDBs
	($v_{tx}=0$), namely
	\begin{align}
		&S_{jj}^{(\mathrm{IB})}(\omega)=\left[\mathrm{Re}~\sigma_{jj}^{\mathrm{(IB)}}(\omega,\mu,t_{x})\right]_{v_{tx}=\mu=0},
	\end{align}
	 which can be explicitly written as
	\begin{align}
		S_{xx}^{(\mathrm{IB})}(\omega) &
		=\mathrm{B}(\frac{3}{4},\frac{3}{2})\frac{\sigma_{0}}{\pi}\frac{\sqrt{2a \omega}}{v_{F}}, \label{Spara} \\
		S_{yy}^{(\mathrm{IB})}(\omega) &
		=\mathrm{B}(\frac{5}{4},\frac{1}{2})\frac{\sigma_{0}}{\pi}\frac{v_{F}}{\sqrt{2a \omega}}, \label{Sperp}
	\end{align}
	where $\sigma_{0}=e^2/(4\hbar)$ (here, $\hbar$ is restored temporarily for clarity), and $\mathrm{B}(p,q)$ is the standard Beta function. It is therefore evident that their geometric mean 
	\begin{align}
		\bar{S}^{(\mathrm{IB})}(\omega)&=\sqrt{S_{xx}^{(\mathrm{IB})}(\omega)\times S_{yy}^{(\mathrm{IB})}(\omega)}=\sqrt{\frac{8}{15\pi}}\sigma_{0}
		\label{Sprods}
	\end{align}
	is a constant, independent of $\omega$. These analytical results---given by Eqs. (\ref{Spara}), (\ref{Sperp}), and (\ref{Sprods})---are identical to those derived for 2D SDBs tilting along the linear dispersion \cite{PRBYan2023} (up to three overall factors due to the redefinition of $S_{jj}^{(\mathrm{IB})}(\omega)$ herein), but differ from those for 2D tilted DBs \cite{PRBTan2022,PRBHou2023}.

	The dimensionless auxiliary function $\Gamma_{jj}^{(\mathrm{IB})}(\omega,\mu,t_{x})$ exhibits a Heaviside-like behavior and can be cast in the unified form
	\begin{align}
		\Gamma_{jj}^{(\mathrm{IB})}(\omega,\mu,t_{x})
		&=\mathcal{C}_{jj}\left(\tilde{\xi}_{+}\right)
		+\mathcal{C}_{jj}\left(\tilde{\xi}_{-}\right),
	\end{align}
	where 
	\begin{align}
		\mathcal{C}_{xx}(z)&=\mathscr{B}\left(z,\frac{3}{4},\frac{3}{2}\right),\\
		\mathcal{C}_{yy}(z)&=\mathscr{B}\left(z,\frac{5}{4},\frac{1}{2}\right),
	\end{align}
	and 
	\begin{align}
		&\tilde{\xi}_{\pm}=
		\begin{cases}
			-1, & \xi_{\pm}\le -1,\\\\
			\xi_{\pm}, &-1<\xi_{\pm}<+1,\\\\
			+1, & +1\le \xi_{\pm},
		\end{cases}
	\end{align}
	with
	\begin{align}
		\mathscr{B}(z,p,q)
		&=\frac{1}{2}\mathrm{sgn}(z)\Theta(1-z^2)\frac{\mathcal{B}(z^4,p,q)}{\mathrm{B}(p,q)},
	\end{align}
	and 
	\begin{align}
		&\xi_{\lambda}= \frac{\omega+(\lambda)2\mu}{2v_{tx}}\sqrt{\frac{2a}{\omega}}.
	\end{align}
	In these definitions, $\mathrm{sgn}(z)$ denotes the sign function, conventionally defined piecewise: $\mathrm{sgn}(z) = -1$ for $z < 0$, $0$ for $z = 0$, and $+1$ for $z > 0$, which satisfies the identity $\mathrm{sgn}(z) = \Theta(z) - \Theta(-z)$ for any real $z$; $\mathcal{B}(\zeta,p,q) = \int_{0}^{\zeta} u^{p-1}(1-u)^{q-1} du$ is the incomplete Beta function for $\zeta \in [0,1]$, and $\mathcal{B}(1,p,q)$ reduces to the standard Beta function $\mathrm{B}(p,q)$. We note that both $\mathscr{B}(z,p,q)$ and $\mathcal{C}_{jj}(z)$ are odd functions of $z$. In addition, the index $\lambda$ in $\xi_\lambda$ labels the contribution from the electron ($\lambda = +$) and hole ($\lambda = -$) pockets, respectively.

\begin{figure*}[htbp]
		\includegraphics[width=18cm]{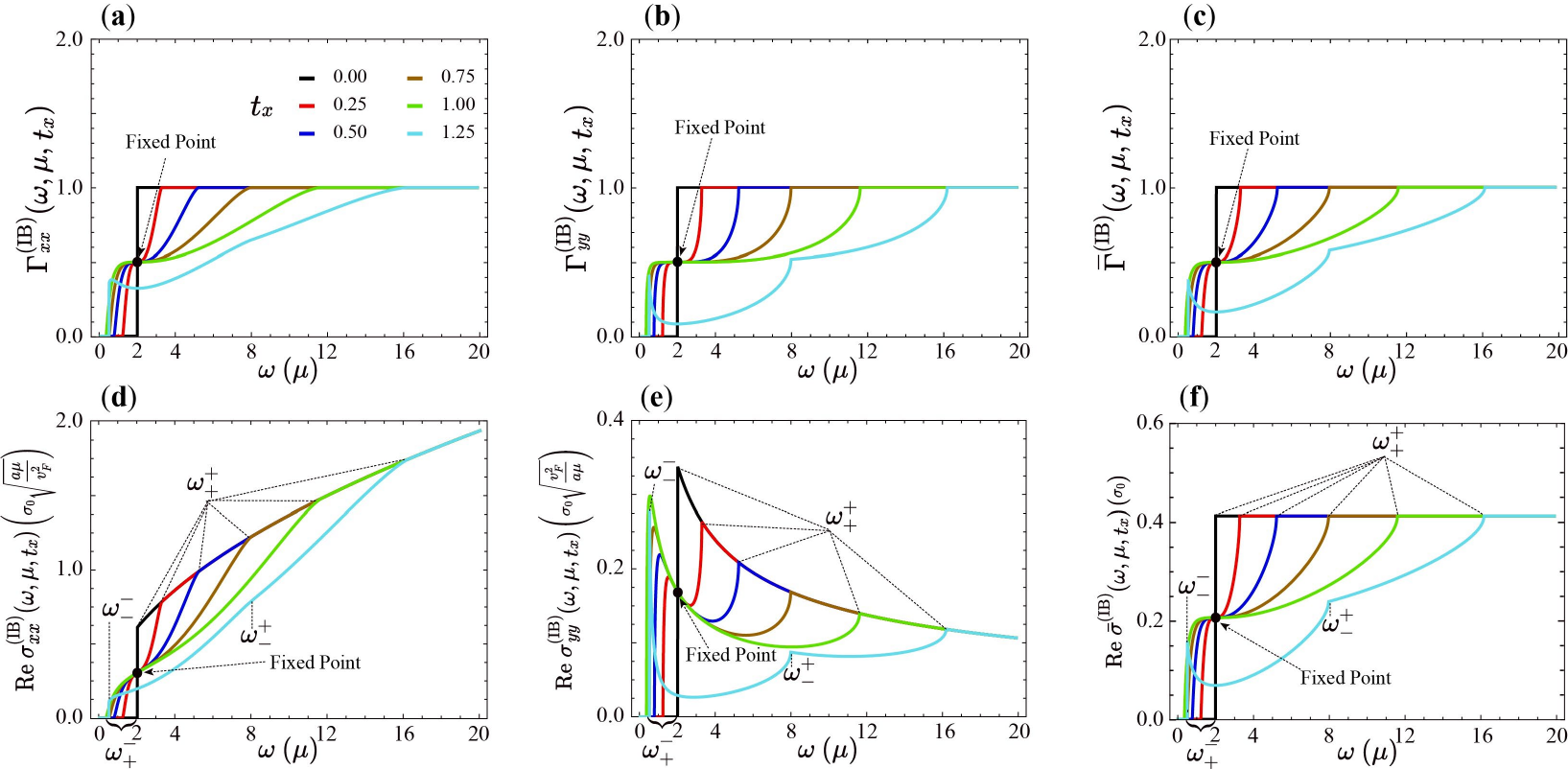}\\
		\caption{Dimensionless auxiliary functions  $\Gamma_{xx}^{(\mathrm{IB})}(\omega,\mu,t_{x})$, $\Gamma_{yy}^{(\mathrm{IB})}(\omega,\mu,t_{x})$ , and their geometric mean $\bar{\Gamma}^{(\mathrm{IB})}(\omega,\mu,t_{x})$ are shown in panels (a)-(c). Interband LOCs $\mathrm{Re}~\sigma_{xx}^{(\mathrm{IB})}(\omega,\mu,t_{x})$, $\mathrm{Re}~\sigma_{yy}^{(\mathrm{IB})}(\omega,\mu,t_{x})$, and their geometric mean $\mathrm{Re}~\bar{\sigma}^{(\mathrm{IB})}(\omega,\mu,t_{x})$ are shown in panels (d)-(f). The fixed points are denoted by the solid black dots. The critical frequencies $\omega_{\lambda}^{\pm}$ in the interband LOCs are indicated in panels (d)-(f). The legends for all panels are shown in panel (a).}
		\label{Figure2}
	\end{figure*}

	It is emphasized that the unified expression is valid for both spatial components ($j=x$ and $j=y$), and all Lifshitz phases: untilted phase ($t_{x}=0$), type-I phase ($0<t_{x}<1$), type-III phase ($t_{x}=1$), and type-II phase ($t_{x}>1$). The explicit expressions for $\Gamma_{jj}^{(\mathrm{IB})}(\omega,\mu,t_{x})$ in different Lifshitz phases in the $n$-doped case ($\mu > 0$) are given in the following. We explicitly show the results regarding interband LOCs in Fig. \ref{Figure2}.
	
	It is noted that there are in general four critical frequencies of interband optical transitions,
	\begin{align}
		\omega_{\lambda}^{\pm} &
		= 2 \mu\left(t_{x}\pm\sqrt{\left|1+\lambda t_{x}^{2}\right|}\right)^2,
	\end{align}
	as illustrated in 
	Fig.~\ref{Figure1} (i)-(l) and Fig.~\ref{Figure2} (d)-(f). For the untilted phase ($t_{x}=0$), the two critical frequencies $\omega_{+}^{\pm}$ contributed by the electron pocket ($\lambda=+$) are degenerate at $\omega = 2\mu$, while other two critical frequencies $\omega_{-}^{\pm}$ contributed by the hole pocket ($\lambda=-$) do not appear. Consequently, the dimensionless auxiliary function simplifies to
	\begin{align}
		\Gamma_{jj}^{(\mathrm{IB})}(\omega,\mu,t_{x}=0)=\Theta(\omega-2\mu),
		\label{untilted}
	\end{align}
	which agrees with previous results \cite{PRBCarbotte2019,PRBYan2023}, and is plotted as solid black lines in Fig.~\ref{Figure2} (a) and (b). In the regime of type-I phase ($0<t_{x}<1$), the two critical frequencies, $\omega_{+}^{+}$ and $\omega_{+}^{-}$, are distinct, and other two critical frequencies $\omega_{-}^{\pm}$ contributed by the hole pocket are still absent. The dimensionless auxiliary function is given by the piecewise expression
	\begin{align}
		&\Gamma_{jj}^{(\mathrm{IB})}(\omega,\mu,0<t_{x}<1)
		\nonumber\\
		&=
		\begin{cases}
			0, & 0<\omega\leq\omega_{+}^{-},\\\\
			\mathcal{C}_{jj}(1) + \mathcal{C}_{jj}\left(\xi_{-}\right), & \omega_{+}^{-}<\omega<\omega_{+}^{+},\\\\
			1, & \omega_{+}^{+}\le\omega.
		\end{cases}
	\end{align}
	which reduces to that for the untilted phase ($t_{x}=0$) in the limit $t_{x}\to 0^{+}$ with $\omega_{+}^{-}=\omega_{+}^{+}$. For the type-II phase ($t_{x}>1$), all four critical frequencies contributed by the electron pocket ($\omega_{+}^{-}$ and $\omega_{+}^{+}$) and hole pocket ($\omega_{-}^{-}$ and $\omega_{-}^{+}$) are distinct. The dimensionless auxiliary function is consequently defined by a more detailed piecewise function
	\begin{align}
		&\Gamma_{jj}^{(\mathrm{IB})}(\omega,\mu,t_{x}>1)
		\nonumber\\&=
		\begin{cases}
			0, & 0<\omega\le\omega_{+}^{-},\\\\
			\mathcal{C}_{jj}(1) + \mathcal{C}_{jj}\left(\xi_{-}\right), & \omega_{+}^{-}<\omega\le\omega_{-}^{-},\\\\
			\mathcal{C}_{jj}\left(\xi_{+}\right) + \mathcal{C}_{jj}\left(\xi_{-}\right), & \omega_{-}^{-}<\omega<\omega_{-}^{+},\\\\
			\mathcal{C}_{jj}(1) + \mathcal{C}{jj}\left(\xi_{-}\right), & \omega_{-}^{+}\le\omega<\omega_{+}^{+},\\\\
			1, & \omega_{+}^{+}\le\omega.
		\end{cases}
	\end{align}
	The dimensionless auxiliary function derived directly at the type-III phase ($t_{x}=1$) is consistent with the limit from either the type-I or type-II phase, demonstrating the continuity of $\Gamma_{jj}^{(\mathrm{IB})}$ with respect to the tilt parameter, namely,
	\begin{align}
		\lim_{ t_{x}\to 1^{\pm} }\Gamma_{jj}^{(\mathrm{IB})}(\omega,\mu,t_{x}) = \Gamma_{jj}^{(\mathrm{IB})}(\omega,\mu,t_{x}=1).
		%\nonumber
	\end{align}
	
	As shown in Fig.~\ref{Figure2}(a)-(c), the dimensionless auxiliary functions $\Gamma_{xx}^{(\mathrm{IB})}(\omega,\mu,t_{x})$, $\Gamma_{yy}^{(\mathrm{IB})}(\omega,\mu,t_{x})$ and their geometric mean $\bar{\Gamma}^{(\mathrm{IB})}(\omega,\mu,t_{x})$ exhibit a Heaviside-like behavior in 2D SDBs tilting along the quadratic dispersion, due to energy conservation and Pauli blocking. Its functional form is segmented into three distinct regions for type-I and type-III phases, but five regions for the type-II phase, which represents a significant departure from the behavior observed in systems tilting along the linear dispersion \cite{PRBYan2023,PRBHou2023,PRBTan2022}. This departure originates mainly from the geometry of Fermi surfaces: one closed Fermi surface for type-I and type-III phases, whereas two closed Fermi surfaces for type-II phase,  as shown in Fig.~\ref{Figure1} (a)-(f). By contrast, there is only one closed Fermi surface for type-I phase, only one open Fermi surface for type-III phase, but two open Fermi surfaces for type-II phase, in 2D SDBs or DBs tilting along the linear dispersion \cite{PRBYan2023,PRBHou2023,PRBTan2022}.

    In the high-frequency regime where $\omega = \Omega \gg \omega_{+}^{+}= 2 \mu\left(t_{x}+\sqrt{\left|1+ t_{x}^{2}\right|}\right)^2 \ge 2\mu$,
    and for any tilt parameter $t_x \ge 0$, the dimensionless auxiliary function approaches unity, namely,
    \begin{align}
    	\Gamma_{jj}^{(\mathrm{IB})}(\Omega,\mu,t_{x}) = 1,
    \end{align}
    as shown in Fig.~\ref{Figure2} (a) and (b).
    This result leads directly to that the asymptotic background value of two interband LOCs
    \begin{align}
    	\mathrm{Re}~\sigma_{jj}^{(\mathrm{IB})}(\omega=\Omega,\mu,t_{x})&
    	= S_{jj}^{(\mathrm{IB})}(\Omega),
    \end{align}
    and their geometric mean
    \begin{align}
    	\mathrm{Re}~\bar{\sigma}^{(\mathrm{IB})}(\omega=\Omega,\mu,t_{x})
    	&
    	=\sqrt{\frac{8}{15\pi}}\sigma_0,
    \end{align}
    which are always independent of the tilt parameter $t_x$ and chemical potential $\mu$ in 2D SDBs tilting along the quadratic dispersion, as demonstrated in Fig. \ref{Figure2} (d)-(f). Consequently, the asymptotic behavior of the interband LOCs can not be used to distinguish different Lifshitz phases experimentally in the present system. On the contrary, when the 2D SDBs and DBs are tilted along the linear dispersion, these asymptotic background values are tilt-independent in the untilted, type-I, and type-III phases but tilt-dependent in the type-II phase, hence can fingerprint different Lifshitz phases \cite{PRBYan2023,PRBTan2022,PRBHou2023}. 
	
	A notable result for 2D SDBs tilting along the quadratic dispersion is that for tilt parameters in the range $0 < t_{x} \le 1$, the two dimensionless auxiliary functions $\Gamma_{xx}^{(\mathrm{IB})}(\omega,\mu,t_{x})$ and $\Gamma_{yy}^{(\mathrm{IB})}(\omega,\mu,t_{x})$ evaluated at $\omega = 2\mu$ remains exactly one half, namely,
	\begin{align}
		\hspace{-0.15cm}
		\Gamma_{jj}^{(\mathrm{IB})}(2\mu,\mu,0 < t_{x} \le 1)& = \frac{\Gamma_{jj}^{(\mathrm{IB})}(2\mu,\mu,t_{x}=0)}{2}
		%\nonumber\\&
		=\frac{1}{2},
		\label{converge}
		\end{align}
		which is always independent of both $\mu$ and $t_{x}$, rendering their geometric mean 
		\begin{align}
		\bar{\Gamma}^{(\mathrm{IB})}(2\mu,\mu,0 < t_{x} \le 1)
		&=\frac{1}{2},
	\end{align}
	    with $\bar{\Gamma}^{(\mathrm{IB})}(\omega,\mu,t_{x})$ the geometric mean of two dimensionless auxiliary functions defined by
	\begin{align} 
		\bar{\Gamma}^{(\mathrm{IB})}(\omega,\mu,t_{x})
		&=\sqrt{\Gamma_{xx}^{(\mathrm{IB})}(\omega,\mu,t_{x})\times\Gamma_{yy}^{(\mathrm{IB})}(\omega,\mu,t_{x})}.
	\end{align}
	Consequently, the interband LOCs $\mathrm{Re}~\sigma_{jj}^{(\mathrm{IB})}(\omega,\mu,t_{x})$  exhibit fixed points at $\omega = 2\mu$, namely,
	\begin{align}
		\mathrm{Re}~\sigma_{jj}^{(\mathrm{IB})}(2\mu,\mu,0 < t_{x}\le 1)
		&= \frac{\mathrm{Re}~\sigma_{jj}^{(\mathrm{IB})}(2\mu,\mu,t_{x}=0)}{2}
		\nonumber\\&
		=\frac{S_{jj}^{(\mathrm{IB})}(2\mu)}{2},
		\label{fixedpoints}
	\end{align}
	 which leads further to
	\begin{align}
		\mathrm{Re}~\bar{\sigma}^{(\mathrm{IB})}(2\mu,\mu,0 < t_{x} \le 1)
		&= \sqrt{\frac{2}{15\pi}} \sigma_0,
	\end{align}
	with $\mathrm{Re}~\bar{\sigma}^{(\mathrm{IB})}(\omega,\mu,t_{x})$ the geometric mean of the two interband LOCs defined as
	\begin{align}  \mathrm{Re}~\bar{\sigma}^{(\mathrm{IB})}(\omega,\mu,t_{x})
		&=\bar{S}^{(\mathrm{IB})}(\omega)\bar{\Gamma}^{(\mathrm{IB})}(\omega,\mu,t_{x}).
	\end{align}
	
	 The fixed points in $\Gamma_{jj}^{(\mathrm{IB})}(\omega,\mu,t_{x})$, $\mathrm{Re}~\sigma_{jj}^{(\mathrm{IB})}(\omega,\mu,t_{x})$, $\bar{\Gamma}^{(\mathrm{IB})}(\omega,\mu,t_{x})$, and $\mathrm{Re}~\bar{\sigma}^{(\mathrm{IB})}(\omega,\mu,t_{x})$ at $\omega = 2\mu$ are indicated by black dots in Fig. \ref{Figure2} (a)-(e). Interestingly, similar fixed points were previously reported in systems tilting along the linear dispersion \cite{PRBYan2023,PRBTan2022,PRBHou2023}. However, a substantial difference is the range of tilt parameter in which the fixed point exists: it occurs for $0 < t_x \le 1$ in the present case, but for $0 < t_y \le 2$ in 2D SDBs and DBs tilting along the linear dispersion.

	The essential features of interband LOCs can be intuitively understood by the JDOS
	\begin{align}
		\mathcal{J} (\omega,\mu,t_{x})
		&=\int^{+\infty}_{-\infty}\frac{dk_x}{2\pi} \int^{+\infty}_{-\infty}\frac{dk_y}{2\pi}
		\nonumber\\&\times
		\sum_{\kappa=\pm}\left\{f\left[\varepsilon_{\kappa}^{-}(k_x,k_y),\mu\right] -f\left[\varepsilon_{\kappa}^{+}(k_x,k_y),\mu\right]\right\}
		\nonumber\\&\times
		\delta\left[\omega+\varepsilon_{+}^{-}(k_x,k_y)-\varepsilon_{+}^{+}(k_x,k_y)\right],
		%\label{eqnJDOS}
	\end{align}
	in which two parts $f\left[\varepsilon_{\kappa}^{-}(k_x,k_y),\mu\right] -f\left[\varepsilon_{\kappa}^{+}(k_x,k_y),\mu\right]$ and $\delta\left[\omega+\varepsilon_{\kappa}^{-}(k_x,k_y)-\varepsilon_{\kappa}^{+}(k_x,k_y)\right]$ in the integrand account for the Pauli exclusion principle and energy conservation obeyed in the process of optical transition, respectively. 

    \begin{figure}[htbp]
    	%\vspace{0.5cm}
    	\includegraphics[width=8cm]{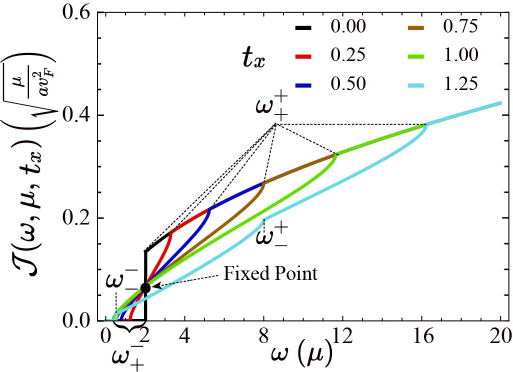}
    	\caption{Joint density of state. The fixed point are denoted by the solid black dots. The critical frequencies $\omega_{\lambda}^{\pm}$ are indicated. Two critical frequencies $\omega_{+}^{\pm}$ contributed by electron pocket ($\lambda=+$) appear in all tilted phases, but other two critical frequencies $\omega_{-}^{\pm}$ contributed by hole pocket ($\lambda=-$) are present only in the type-II phase.}
    	\label{Figure3}
    \end{figure}
	
	From analytical expressions listed in Appendix~ \ref{App3} and curves in Fig.~\ref{Figure3}, one finds that the JDOS $\mathcal{J}(\omega,\mu,t_{x})$ is always equal to zero if $0<\omega\le\omega_{+}^{-}$ because the interband optical transition is forbidden by Pauli exclusion principle. For arbitrary tilt parameter, the JDOS start to smoothly increase from $\omega=\omega_{+}^{-}$ to $\omega=\omega_{+}^{+}$ due to the contribution of electron pocket ($\lambda=+$).  Interestingly, the discontinuity in the first derivative of JDOS at $\omega=\omega_{+}^{+}$ results in a kink of $\mathcal{J}(\omega,\mu,t_{x})$, responsible for the sharp feature of the interband LOCs at $\omega=\omega_{+}^{+}$, as shown in Fig.~\ref{Figure2} (b) and (e). For $t_x>1$, other two critical frequencies $\omega_{-}^{-}$ and $\omega_{-}^{+}$ emerge due to the contribution of hole pocket ($\lambda=-$). The JDOS at zero temperature can qualitatively explain the behaviors of interband LOCs in 2D SDBs tilting along the quadratic dispersion.
	
	\begin{figure*}[htbp]
		%\vspace{0.5cm}
		\includegraphics[width=13cm]{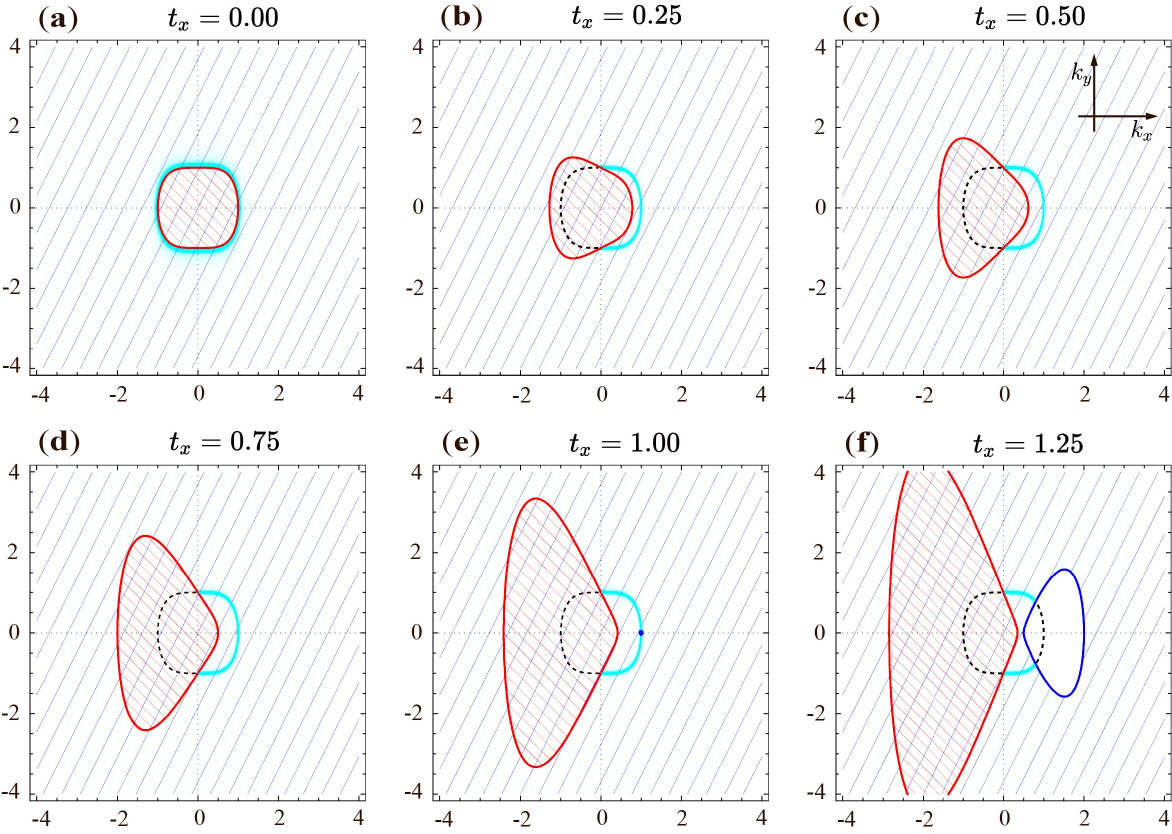}
		\caption{The states involved in the interband optical transition with $\omega=2\mu$ are gathered around a cyan stripe, and the Fermi surfaces are denoted by the red and blue lines. The location of states occupying below them are meshed with light-blue and light-red slashes, and the crossed regions represent that the interband optical transition therein is forbidden by Pauli exclusion principle, rendering the corresponding states represented by the black dashed lines do not contribute to the interband optical transition.}
		\label{Figure4}
	\end{figure*}

	The fixed point of JDOS at $\omega=2\mu$ in Fig.~\ref{Figure3} accounts for the fixed points of $\Gamma_{jj}^{(\mathrm{IB})}(\omega,\mu,t_{x})$, $\mathrm{Re}~\sigma_{jj}^{(\mathrm{IB})}(\omega,\mu,t_{x})$, $\bar{\Gamma}^{(\mathrm{IB})}(\omega,\mu,t_{x})$, and $\mathrm{Re}~\bar{\sigma}^{(\mathrm{IB})}(\omega,\mu,t_{x})$ in Fig.~\ref{Figure2}, which can be confirmed by a strong relation
	\begin{align}
    \hspace{-0.25cm}
		\mathcal{J}(2\mu+0^{+},\mu,0<t_{x}\le 1)=\frac{\mathcal{J}(2\mu+0^{+},\mu,t_{x}=0)}{2},
		\label{strongrelation}
	\end{align}
	whose detailed proof is found in Appendix \ref{App4}. To intuitively visualize the JDOS at $\omega=2\mu$, we identify the states involved in the interband optical transition by cyan stripes as shown in Fig.~\ref{Figure4}, where the red and blue lines denote the contours of Fermi surface. The states involved in the interband optical transition with $\varepsilon_\kappa^{+}(k_x,k_y)-\varepsilon_\kappa^{-}(k_x,k_y)=2\mu$ gather around the same cyan strip. As evidently shown in Fig.~\ref{Figure4}(a)-(e), the states marked in cyan strip for $0<t_{x}\le 1$ are exactly half of those for $t_{x}=0$, illuminating the relation in Eq. \ref{strongrelation} and dictating the relations in Eqs.~(\ref{untilted}), (\ref{converge}), and (\ref{fixedpoints}).
    When $t_x>1$, part of the states marked in cyan strip are forbidden by Pauli principle, leading to $\Gamma_{jj}^{(\mathrm{IB})}(2\mu,\mu,t_{x}>1)<\frac{1}{2}$. As a result, the corresponding fixed points of $\Gamma_{jj}^{(\mathrm{IB})}(\omega,\mu,t_{x})$, $\mathrm{Re}~\sigma_{jj}^{(\mathrm{IB})}(\omega,\mu,t_{x})$, $\bar{\Gamma}^{(\mathrm{IB})}(\omega,\mu,t_{x})$, and $\mathrm{Re}~\bar{\sigma}^{(\mathrm{IB})}(\omega,\mu,t_{x})$ appear when $0<t_{x}\le 1$, regardless of the value of chemical potential $\mu$ and tilt parameter $t_{x}$, as shown in Fig. \ref{Figure2}.

    %%%%%%%%%%%%%%%%%%%%%%%%%%%%%%%	
	\section{Comparisons and Conclusions\label{Sec: Comparisons and Conclusions}}
	%%%%%%%%%%%%%%%%%%%%%%%%%%%%%%%
	
\begin{figure*}[htbp]
        \includegraphics[width=18cm]{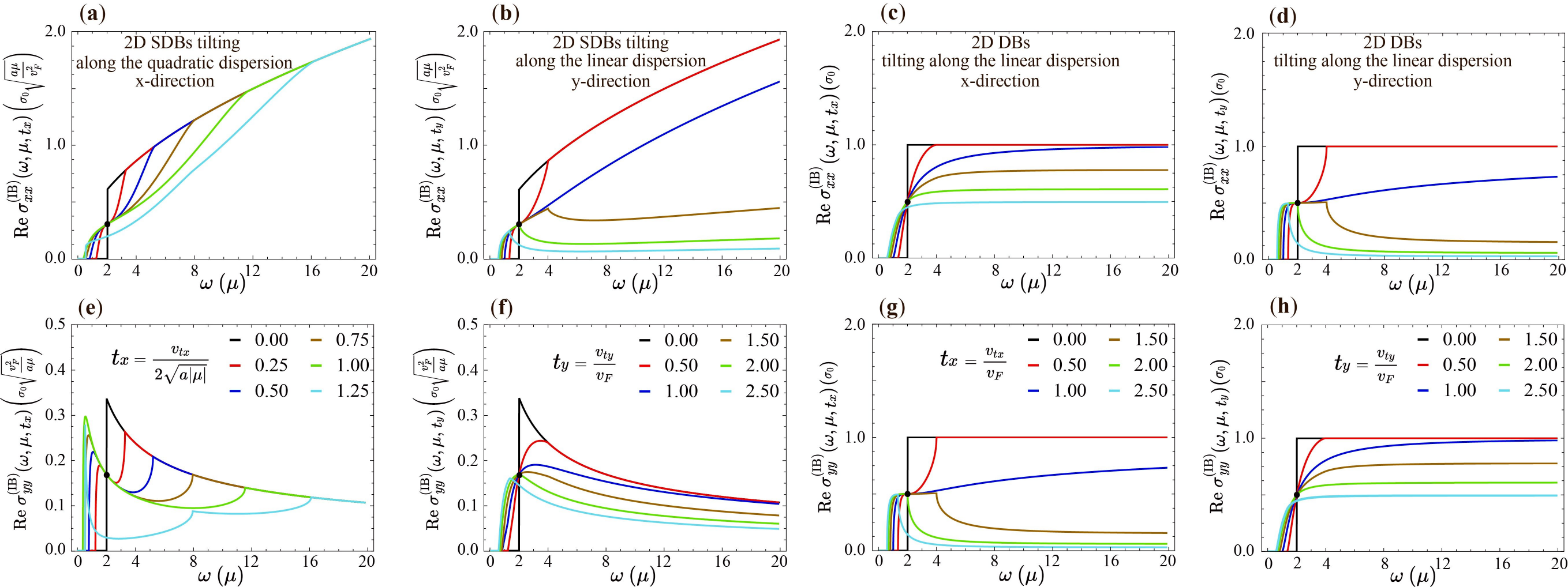}
		\caption{Interband LOCs  $\mathrm{Re}~\sigma_{jj}^{(\mathrm{IB})}(\omega,\mu,t_{i})$ with $i,j$ covering both $x$ and $y$. The analytical results adopted to plot this figure are based on the present work and Refs. \cite{PRBTan2022,PRBHou2023,PRBYan2023}. In panels (c), (d), (g) and (h), $\mathrm{Re}~\sigma_{xx}^{(\mathrm{IB})}(\omega,\mu,t_{x})=\mathrm{Re}~\sigma_{yy}^{(\mathrm{IB})}(\omega,\mu,t_{y})$ and $\mathrm{Re}~\sigma_{xx}^{(\mathrm{IB})}(\omega,\mu,t_{y})=\mathrm{Re}~\sigma_{yy}^{(\mathrm{IB})}(\omega,\mu,t_{x})$. Each column represents a distinct group of panels.}
		\label{Fig.comparison}
	\end{figure*}

    The interband LOCs are governed by the combined effects of the power-law scalings and dimensionless auxiliary functions. We make a comprehensive comparison among the interband LOCs in different kinds of 2D tilted SDBs and DBs in Table \ref{Table:comparison} and Fig. \ref{Fig.comparison} for  
   highlighting their qualitative differences and similarities. As shown in Table \ref{Table:comparison}, $S_{jj}^{(\mathrm{IB})}(\omega)$ always exhibits spatial-direction-anisotropic power-law scalings of $\omega$ in 2D tilted SDBs, unlike the spatial-direction-isotropic one in 2D tilted DBs. Additionally, the power-law scalings are independent of the tilting direction, namely, which spatial direction the 2D SDBs or 2D DBs are tilted along. 

   As demonstrated in Fig.~\ref{Fig.comparison}, the interband LOCs $\mathrm{Re}~\sigma_{jj}^{(\mathrm{IB})}(\omega,\mu,t_{i})$ in 2D tilted SDBs depend remarkably on both the tilting direction and the spatial direction. The dependence arises primarily from the spatial-direction-dependent power-law scalings, and secondarily from the dimensionless auxiliary functions that are influenced by both tilting direction and spatial direction, especially in the type-II phase. Consequently, $\mathrm{Re}~\sigma_{jj}^{(\mathrm{IB})}(\omega,\mu,t_{i})$ in 2D tilted SDBs along different spatial dispersions differ significantly from each other. By contrast, in 2D tilted DBs, the interband LOCs share exactly the same dependence of $\omega$ with the dimensionless auxiliary functions because of spatial-direction-isotropic constant scaling $\sigma_{0}$. Further, both the interband LOCs and the dimensionless auxiliary functions in 2D tilted DBs rely only on the tilting direction, regardless of the spatial direction. Therefore, 2D tilted SDBs and DBs exhibit notable differences.

\begin{table*}[htbp]
    \begin{tabular}{c|c|c|c|c}
        \hline\hline
        \footnotesize{Energy bands} & \multicolumn{2}{c|}{\footnotesize{2D tilted SDBs}} & \multicolumn{2}{c}{\footnotesize{2D tilted DBs}} \\ \hline
        \footnotesize{Tilting direction} & \footnotesize{tilting along x-direction} & \footnotesize{tilting along y-direction} & \footnotesize{tilting along x-direction} & \footnotesize{tilting along y-direction} \\ \hline
        \footnotesize{Hamiltonian} & $\footnotesize{\kappa v_{tx} k_x\tau_0 + a k_x^2 \tau_1+ v_{F} k_y\tau_2}$ & $\footnotesize{\kappa v_{ty} k_y\tau_0 + a k_x^2 \tau_1+ v_{F} k_y\tau_2}$ & $\footnotesize{\kappa v_{tx} k_x\tau_0 + v_{F} (k_x \tau_1+ k_y\tau_2)}$ & $\footnotesize{\kappa v_{ty} k_y\tau_0 + v_{F} (k_x \tau_1+ k_y\tau_2)}$ \\ \hline
        \footnotesize{Tilt parameter $t_{i}$} & $\footnotesize{t_{x}=\frac{v_{tx}}{2\sqrt{a|\mu|}}}$ & $\footnotesize{t_{y}=\frac{v_{ty}}{v_{F}}}$ & $\footnotesize{t_{x}=\frac{v_{tx}}{v_{F}}}$ & $\footnotesize{t_{y}=\frac{v_{ty}}{v_{F}}}$ \\ \hline
        \footnotesize{$S_{xx}^{(\mathrm{IB})}(\omega)$} & $\footnotesize{\mathrm{B}(\frac{3}{4},\frac{3}{2})\frac{\sigma_{0}}{\pi}\frac{\sqrt{2a \omega}}{v_{F}}}$ & $\footnotesize{\mathrm{B}(\frac{3}{4},\frac{3}{2})\frac{\sigma_{0}}{\pi}\frac{\sqrt{2a \omega}}{v_{F}}}$ & $\footnotesize{\sigma_{0}}$ & $\footnotesize{\sigma_{0}}$ \\ \hline
        \footnotesize{$S_{yy}^{(\mathrm{IB})}(\omega)$} & $\footnotesize{\mathrm{B}(\frac{5}{4},\frac{1}{2})\frac{\sigma_{0}}{\pi}\frac{v_{F}}{\sqrt{2a \omega}}}$ & $\footnotesize{\mathrm{B}(\frac{5}{4},\frac{1}{2})\frac{\sigma_{0}}{\pi}\frac{v_{F}}{\sqrt{2a \omega}}}$ & $\footnotesize{\sigma_{0}}$ & $\footnotesize{\sigma_{0}}$ \\ \hline
        \footnotesize{$\mathrm{Re}~\sigma_{xx}^{(\mathrm{IB})}(\omega,\mu,t_{i})$} & \footnotesize{Fig. \ref{Fig.comparison}(a)} & \footnotesize{Fig. \ref{Fig.comparison}(b)} & \footnotesize{Fig. \ref{Fig.comparison}(c)} & \footnotesize{Fig. \ref{Fig.comparison}(d)} \\ \hline
        \footnotesize{$\mathrm{Re}~\sigma_{yy}^{(\mathrm{IB})}(\omega,\mu,t_{i})$} & \footnotesize{Fig. \ref{Fig.comparison}(e)} & \footnotesize{Fig. \ref{Fig.comparison}(f)} & \footnotesize{Fig. \ref{Fig.comparison}(g)} & \footnotesize{Fig. \ref{Fig.comparison}(h)} \\ \hline\hline
    \end{tabular}
    \caption{Comparisons among the model Hamiltonians, tilt parameters, power-law scalings, interband LOCs in 2D tilted SDBs and DBs. The comparisons are based on the analytical resutls in the present work and Refs. \cite{PRBTan2022,PRBHou2023,PRBYan2023}.}
    \label{Table:comparison}
\end{table*}

	The fixed point in the LOCs $\mathrm{Re}~\sigma_{jj}^{(\mathrm{IB})}(\omega,\mu,t_{i})$ appears at $\omega=2\mu$ when 2D SDBs or DBs are tilted along either the linear dispersion or the quadratic dispersion. However, the ranges of tilt parameter for the presence of fixed point are different: $0<t_{x}\le 1$ when the 2D SDBs are tilted along the quadratic dispersion, while $0<t_{y}\le 2$ when the 2D SDBs (and 2D DBs) are tilted along the linear dispersion. In accompany with the findings regarding the fixed point in our previous papers \cite{PRBTan2022,PRBHou2023,PRBYan2023}, we claim here that the interband LOCs in tilted bands host a fixed point for a suitable range of tilt parameter.
	
	In conclusion, we proposed a definition of tilt parameter to characterize the Lifshitz phases in 2D SDBs tilting along the quadratic dispersion, and investigated the highly
	anisotropic interband optical conductivities for distinct Lifshitz phases therein. We found the interband LOCs in 2D SDBs along the quadratic dispersion exhibit notable differences from those in 2D SDBs and DBs along the linear dispersion. We claimed that a robust fixed point appears at $\omega=2\mu$ in the interband optical conductivity of tilted bands in a specific range of tilt parameter. Our theoretical predictions can not only be used to fingerprint different kind of 2D tilted SDBs and DBs in optical measurements, but also guide future experimental studies of anisotropic band engineering and tilt-dependent phenomena. %\LLcmt{What kind of experimental studies other than the optical conductivity measurement?}
	
	%%%%%%%%%%%%%%%%%%%%%%%%%%%%%%%
	\section*{ACKNOWLEDGEMENTS\label{Sec:acknowledgements}}
	%%%%%%%%%%%%%%%%%%%%%%%%%%%%%%%
	
	This work is partially supported by the National Natural Science Foundation of China under Grants No. 11547200 and No. 12204329, the Natural Science Foundation of Jiangsu Province under Grant No. BK20241929, and the Research Institute of Intelligent Manufacturing Industry Technology of Sichuan Arts and Science University.

	%%%%%%%%%%%%%%%%%%%%%%%%%%%%%%%
	\section*{DATA AVAILABILITY}
	%%%%%%%%%%%%%%%%%%%%%%%%%%%%%%%

	The data that support the findings of this article are not
	publicly available upon publication. The data are available from the authors upon reasonable request.

	\appendix
	\allowdisplaybreaks[4]
	\begin{widetext}
		
		%%%%%%%%%%%%%%%%%%%%%%%%%%%%%%%%%%%%%%%%%%%%%%%
		\section{Detailed calculation of joint density of state \label{App3}}
		%%%%%%%%%%%%%%%%%%%%%%%%%%%%%%%%%%%%%%%%%%%%%%%

		The interband LOCs are generally determined by the interband optical transition, involved to which the number of states can be calculated by the JDOS via
		\begin{align}
			\mathcal{J} (\omega,\mu,t_{x})&
			=\int^{+\infty}_{-\infty}\frac{dk_x}{2\pi} \int^{+\infty}_{-\infty}\frac{dk_y}{2\pi}
			\sum_{\kappa=\pm}\left\{f\left[\varepsilon_{\kappa}^{-}(k_x,k_y),\mu\right] -f\left[\varepsilon_{\kappa}^{+}(k_x,k_y),\mu\right]\right\}
			\delta\left[\omega+\varepsilon_{+}^{-}(k_x,k_y)-\varepsilon_{+}^{+}(k_x,k_y)\right],
		\end{align}
		where $\left\{f\left[\varepsilon_{\kappa}^{-}(k_x,k_y),\mu\right] -f\left[\varepsilon_{\kappa}^{+}(k_x,k_y),\mu\right]\right\}$ and $\delta\left[\omega+\varepsilon_{\kappa}^{-}(k_x,k_y)-\varepsilon_{\kappa}^{+}(k_x,k_y)\right]$ account for the Pauli exclusion principle and energy conservation obeyed in the process of optical transition, respectively. 
		
		At zero temperature, the JDOS reduces to 
		\begin{align}
			&\mathcal{J}\left(\omega,\mu,t_{x}\right)
			=\int_{-\infty}^{+\infty}\frac{dk_x}{2\pi}\int_{-\infty}^{+\infty}\frac{dk_y}{2\pi}\sum_{\kappa=\pm}
			\left\{\Theta\left[\mu-\varepsilon_\kappa^-\left(k_x,k_y\right)\right]- \Theta\left[\mu-\varepsilon_\kappa^+\left(k_x,k_y\right)\right]\right\}
			\delta\left[\omega-2\sqrt{\left(ak_x^2\right)^2+v_F^2k_y^2}\right]
			.
		\end{align}
		 where $\Theta(z)$ is defined piecewise as: $\Theta(z)=0$ for $z<0$, and $\Theta(z)=1$ for $z\ge 0$, which satisfies the relation $\Theta(z)+\Theta(-z)=1$. This function satisfies the identity $\Theta(z)+\Theta(-z)=1$. Note that the ambiguity at $\Theta(0)$ can be resolved by defining
         	\begin{align}
			\lim_{z\to 0^{+}}\Theta(z)&\equiv 1,\\		\lim_{z\to 0^{+}}\Theta(-z)&=\lim_{z\to - 0^{+}}\Theta(z)\equiv 0.
		\end{align}

    By utilizing the following relations
         \begin{align}
	       \varepsilon_{-\kappa}^{-\lambda}(k_x,k_y)&=-\varepsilon_{+\kappa}^{+\lambda}(k_x,k_y),\\
	       \varepsilon_{-\kappa}^{+\lambda}(k_x,k_y)&=-\varepsilon_{+\kappa}^{-\lambda}(k_x,k_y),\\
	       \varepsilon_{+\kappa}^{+\lambda}(-k_x,k_y)&=+\varepsilon_{-\kappa}^{+\lambda}(k_x,k_y),\\
	       \varepsilon_{+\kappa}^{+\lambda}(-k_x,k_y)&=-\varepsilon_{+\kappa}^{-\lambda}(k_x,k_y),
     \end{align}
     in the symmetric region $k_{x} \in (-\infty,+\infty)$,
     and introducing two new variables $\xi$ and $\psi$ via
     \begin{align}
			k_x&=\sqrt{\frac{\omega}{2a}}\xi,\\		k_y&=\frac{\omega}{2 v_{F}}\psi,
		\end{align}
    we have
     \begin{align}
			&\mathcal{J}\left(\omega,\mu,t_{x}\right)
			=\frac{4}{\sqrt{a}v_F}\frac{1}{4\pi^{2}} \sqrt{\frac{\omega}{2}} \frac{\omega}{2}\int_{-\infty}^{+\infty}d \xi\int_{0}^{+\infty}d \psi
			~\delta\left[\omega-\omega\sqrt{\xi^4+\psi^2}\right]
			\nonumber\\&\hspace{2cm}
			\left\{\Theta\left[\mu-\left(v_{tx}\sqrt{\frac{\omega}{2a}}\xi-\frac{\omega}{2} \sqrt{\xi^4+\psi^2}\right)\right]- \Theta\left[\mu-\left(v_{tx}\sqrt{\frac{\omega}{2a}}\xi+\frac{\omega}{2} \sqrt{\xi^4+\psi^2}\right)\right]\right\}
			\nonumber\\&
			=\frac{4}{\sqrt{a}v_F}\frac{1}{4\pi^{2}}\sqrt{\frac{\omega}{2}} \frac{\omega}{2}\frac{1}{\omega}\frac{1}{4}\int_{-\infty}^{+\infty}d \xi
			~\frac{4}{\sqrt{1-\xi^4}}\Theta\left(1-\xi^{2}\right)
			~\left\{\Theta\left[\mu-v_{tx}\sqrt{\frac{\omega}{2a}}\xi+\frac{\omega}{2}\right]- \Theta\left[\mu-v_{tx}\sqrt{\frac{\omega}{2a}}\xi-\frac{\omega}{2}\right]\right\}.
		\end{align}

    We can express the JDOS in the following form,
        \begin{align}
			\mathcal{J}\left(\omega,\mu,t_{x}\right)
			&=S_{J}(\omega)\Gamma_{J}\left(\omega,\mu,t_{x}\right),
		\end{align}
        where
		\begin{align}
			S_{J}(\omega)&=\left[\mathcal{J}\left(\omega,\mu,t_{x}\right)\right]_{v_{tx}=\mu=0}
			=\frac{4}{\sqrt{a}v_F}\frac{1}{4\pi^{2}}\sqrt{\frac{\omega}{2}} \frac{\omega}{2}\frac{1}{\omega}\frac{1}{4}\int_{-\infty}^{+\infty}d \xi
			~\frac{4}{\sqrt{1-\xi^4}}\Theta\left(1-\xi^{2}\right)
			~\left\{\Theta\left[+\frac{\omega}{2}\right]- \Theta\left[-\frac{\omega}{2}\right]\right\}
			\nonumber\\&
			=\frac{1}{8\pi^2v_F}\sqrt{\frac{\omega}{2a}}\int_{-\infty}^{+\infty}d \xi
			~\frac{4}{\sqrt{1-\xi^4}}\Theta\left(1-\xi^{2}\right)
			=\frac{\mathrm{B}\left(\frac{1}{4},\frac{1}{2}\right)}{4\pi^2v_F}\sqrt{\frac{\omega}{2a}}
		\end{align}
		is a dimensional auxiliary function, and $\Gamma_{J}\left(\omega,\mu,t_{x}\right)$ denotes a dimensionless auxiliary function
		\begin{align}
		\Gamma_{J}\left(\omega,\mu,t_{x}\right)
			&=\frac{1}{2\mathrm{B}\left(\frac{1}{4},\frac{1}{2}\right)}\int_{-\infty}^{+\infty}d \xi~4\left(1-\xi^4\right)^{-\frac{1}{2}}~\Theta\left(1-\xi^{2}\right)
			~\left\{\Theta\left[\mu-v_{tx}\sqrt{\frac{\omega}{2a}}\xi+\frac{\omega}{2}\right]- \Theta\left[\mu-v_{tx}\sqrt{\frac{\omega}{2a}}\xi-\frac{\omega}{2}\right]\right\}
			\nonumber\\
			&=\frac{1}{2\mathrm{B}\left(\frac{1}{4},\frac{1}{2}\right)}\int_{0}^{+1}d \xi~4\left(1-\xi^4\right)^{-\frac{1}{2}}
			~\left\{\mathrm{sgn}\left(\frac{\omega}{2}+\mu\right)\Theta\left[\left(\frac{\omega}{2}+\mu\right)^2-\frac{v_{tx}^2}{2a}\omega\xi^{2}\right]
			\right.\nonumber\\&\left.\hspace{5.3cm}
			+\mathrm{sgn}\left(\frac{\omega}{2}-\mu\right)\Theta\left[\left(\frac{\omega}{2}-\mu\right)^2-\frac{v_{tx}^2}{2a}\omega\xi^2\right]\right\}.
		\end{align}

		Next, we list $\Gamma_{J}\left(\omega,\mu,t_{x}\right)$ for untilted and tilted cases. In the untilted phase ($t_{x}=0$) and doped case ($\mu>0$), for $\omega>0$, we have 
		\begin{align}
			\Gamma_{J}\left(\omega,\mu,t_{x}=0\right)
			&=\frac{1}{2\mathrm{B}\left(\frac{1}{4},\frac{1}{2}\right)}\int_{0}^{+1}d \xi~4\left(1-\xi^4\right)^{-\frac{1}{2}}
			~\left\{\mathrm{sgn}\left(\frac{\omega}{2}+\mu\right)\Theta\left[\left(\frac{\omega}{2}+\mu\right)^2\right]
			+\mathrm{sgn}\left(\frac{\omega}{2}-\mu\right)\Theta\left[\left(\frac{\omega}{2}-\mu\right)^2\right]\right\}
			\nonumber\\&
			=\frac{1}{2\mathrm{B}\left(\frac{1}{4},\frac{1}{2}\right)}
			\left\{
			\int_{0}^{+1}d u~u^{\frac{1}{4}-1}(1-u)^{\frac{1}{2}-1}
			+\mathrm{sgn}\left(\frac{\omega}{2}-\mu\right)\Theta\left[\left(\frac{\omega}{2}-\mu\right)^2\right]\int_{0}^{+1}d u~u^{\frac{1}{4}-1}(1-u)^{\frac{1}{2}-1}\right\}
			\nonumber\\&
			=\begin{cases}
				0,    &0<\omega<2\mu,\\\\
				1,  &2\mu< \omega,
			\end{cases}
		\end{align}
		with $u=\xi^4$.

		In the tilted phase ($t_{x}>0$) and doped case ($\mu>0$), for $\omega>0$, we have 
		\begin{align}
			\Gamma_{J}(\omega,\mu,t_{x})
			&=\frac{1}{2\mathrm{B}\left(\frac{1}{4},\frac{1}{2}\right)}\int_{0}^{+1}d \xi
			~4\left(1-\xi^4\right)^{-\frac{1}{2}}
			~\left\{\mathrm{sgn}\left(\frac{\omega}{2}+\mu\right)\Theta\left[\left(\frac{\omega}{2}+\mu\right)^2-v_{tx}^2\frac{\omega}{2a}^2\right]
			\right.\nonumber\\&\left.\hspace{5.3cm}
			+\mathrm{sgn}\left(\frac{\omega}{2}-\mu\right)\Theta\left[\left(\frac{\omega}{2}-\mu\right)^2-v_{tx}^2\frac{\omega}{2a}\xi^2\right]\right\}
			\nonumber\\&
			=\frac{1}{2\mathrm{B}\left(\frac{1}{4},\frac{1}{2}\right)}\int_{0}^{+1}d \xi~4\left(1-\xi^4\right)^{-\frac{1}{2}}
			~\left\{
			\mathrm{sgn}(\xi_{+})\Theta\left(\xi_{+}^{2}-\xi^{2}\right)
			+\mathrm{sgn}(\xi_{-})\Theta\left(\xi_{-}^{2}-\xi^{2}\right)\right\}
			\nonumber\\&
			=\frac{1}{2\mathrm{B}\left(\frac{1}{4},\frac{1}{2}\right)}\sum_{\lambda=\pm}\mathrm{sgn}(\xi_{\lambda})\int_{0}^{+1}d \xi~4\left(1-\xi^4\right)^{-\frac{1}{2}}~
			\Theta\left(\xi_{\lambda}^{2}-\xi^{2}\right)
			\nonumber\\
			&=\frac{1}{2\mathrm{B}\left(\frac{1}{4},\frac{1}{2}\right)}\sum_{\lambda=\pm}~\mathrm{sgn}(\xi_{\lambda})\left\{\Theta\left[\xi_{\lambda}^2-1\right]\int_{0}^{+1}du~u^{\frac{1}{4}-1}(1-u)^{\frac{1}{2}-1}
			+\Theta\left[1-\xi_{\lambda}^4\right]\int_{0}^{\xi_{\lambda}^2}du~u^{\frac{1}{4}-1}(1-u)^{\frac{1}{2}-1}
			\right\}
			\nonumber\\&
			=\sum_{\lambda=\pm}~\mathscr{B}\left(\tilde{\xi}_{\lambda},\frac{1}{4},\frac{1}{2}\right)
			=\mathscr{B}\left(\tilde{\xi}_{+},\frac{1}{4},\frac{1}{2}\right)
			+\mathscr{B}\left(\tilde{\xi}_{-},\frac{1}{4},\frac{1}{2}\right).
		\end{align}

		Consequently, we have
		\begin{align}
			\Gamma_{J} (\omega,\mu,0<t_{x}<1) 
			=\begin{cases}
				0,    &0<\omega\le\omega_{+}^{-},\\\\
				\mathscr{B}\left(1,\frac{1}{4},\frac{1}{2}\right)
				+\mathscr{B}\left(\xi_{-},\frac{1}{4},\frac{1}{2}\right),   &\omega_{+}^{-}<\omega<\omega_{+}^{+},\\\\
				1, &\omega_{+}^{+}\le \omega,
			\end{cases}
		\end{align}
		\begin{align}
			\Gamma_{J} (\omega,\mu,t_{x}>1) 
			= \begin{cases}
				0,    &0<\omega\le\omega_{+}^{-},\\\\
				\mathscr{B}\left(1,\frac{1}{4},\frac{1}{2}\right)
				+\mathscr{B}\left(\xi_-,\frac{1}{4},\frac{1}{2}\right),   
				&\omega_{+}^{-}<\omega\le\omega_{-}^{-},\\\\
				\mathscr{B}\left(\xi_+,\frac{1}{4},\frac{1}{2}\right)
				+\mathscr{B}\left(\xi_-,\frac{1}{4},\frac{1}{2}\right), 
				&\omega_{-}^{-}<\omega<\omega_{-}^{+},\\\\
				\mathscr{B}\left(1,\frac{1}{4},\frac{1}{2}\right)
				+\mathscr{B}\left(\xi_-,\frac{1}{4},\frac{1}{2}\right),  &\omega_{-}^{+}\le\omega<\omega_{+}^{+},\\\\
				1,
				&\omega_{+}^{+}\le\omega,
			\end{cases}
		\end{align}
		and
		\begin{align}
			\Gamma_{J} (\omega,\mu,t_{x}=1) 
			= \begin{cases}
				0,    &0<\omega\le\omega_{+}^{-},\\\\
				\mathscr{B}\left(1,\frac{1}{4},\frac{1}{2}\right)
				+\mathscr{B}\left(\xi_-,\frac{1}{4},\frac{1}{2}\right),   &\omega_{+}^{-}<\omega<\omega_{+}^{+},\\\\
				1,  &\omega_{+}^{+}\le\omega.
			\end{cases}
		\end{align}
		with $\mathscr{B}\left(1,\frac{1}{4},\frac{1}{2}\right)=\frac{1}{2}$.
		
		%%%%%%%%%%%%%%%%%%%%%%%%%%%%%%%%%%%%%%%%%%%%%%%
		\section{The fixed point at $\omega=2\mu$ \label{App4}}
		%%%%%%%%%%%%%%%%%%%%%%%%%%%%%%%%%%%%%%%%%%%%%%%

		By definition, the JDOS at $\omega=2\mu\gg 0^{+}$ takes
		\begin{align}
			\mathcal{J}\left(2\mu,\mu,t_{x}\right)
			&=S_{J}(2\mu)~\Gamma_{J} (2\mu,\mu,t_{x}),
		\end{align}
		where
		\begin{align}
			S_{J}(2\mu)&=\frac{\mathrm{B}\left(\frac{1}{4},\frac{1}{2}\right)}{4\pi^2v_F}\sqrt{\frac{2\mu}{2a}}
           =\frac{\mathrm{B}\left(\frac{1}{4},\frac{1}{2}\right)}{4\pi^2v_F}\sqrt{\frac{\mu}{a}}.
		\end{align}
		
		It is noted that $\Gamma_{J}\left(\omega,\mu,t_{x}\right)$ could be discontinuous at $\omega=2\mu\gg 0^{+}$, thus we take the limit from above
		\begin{align}
			&\lim_{\omega\to 2\mu+0^{+}}\Gamma_{J}\left(\omega,\mu,t_{x}\right)
			\nonumber\\&
			=\frac{1}{2\mathrm{B}\left(\frac{1}{4},\frac{1}{2}\right)}\int_{0}^{+1}d \xi~4\left(1-\xi^4\right)^{-\frac{1}{2}}
			~\left\{\mathrm{sgn}\left(\frac{2\mu+0^{+}}{2}+\mu\right)\Theta\left[\left(\frac{2\mu+0^{+}}{2}+\mu\right)^2-\frac{v_{tx}^2}{2a}(2\mu+0^{+})\xi^{2}\right]
			\right.\nonumber\\&\hspace{5.35cm}\left.
			+\mathrm{sgn}\left(\frac{2\mu+0^{+}}{2}-\mu\right)\Theta\left[\left(\frac{2\mu+0^{+}}{2}-\mu\right)^2-\frac{v_{tx}^2}{2a}(2\mu+0^{+})\xi^2\right]\right\}
			\nonumber\\
			&=\frac{1}{2\mathrm{B}\left(\frac{1}{4},\frac{1}{2}\right)}\int_{0}^{+1}d \xi~4\left(1-\xi^4\right)^{-\frac{1}{2}}
			~\left\{\mathrm{sgn}\left(2\mu+0^{+}\right)\Theta\left[\left(2\mu+0^{+}\right)^2-\frac{v_{tx}^2}{2a}(2\mu+0^{+})\xi^{2}\right]
			\right.\nonumber\\&\hspace{5.35cm}\left.
			+\mathrm{sgn}\left(0^{+}\right)\Theta\left[\left(0^{+}\right)^2-\frac{v_{tx}^2}{2a}(2\mu+0^{+})\xi^2\right]\right\}
			\nonumber\\
			&=\frac{1}{2\mathrm{B}\left(\frac{1}{4},\frac{1}{2}\right)}\int_{0}^{+1}d \xi~4\left(1-\xi^4\right)^{-\frac{1}{2}}
			~\left\{\Theta\left[\left(2\mu\right)^2-\frac{v_{tx}^2}{2a}(2\mu)\xi^{2}\right]
			+\Theta\left[0^{+}-\frac{v_{tx}^2}{2a}(2\mu)\xi^2\right]\right\}
			\nonumber\\
			&=\frac{1}{2\mathrm{B}\left(\frac{1}{4},\frac{1}{2}\right)}\int_{0}^{+1}d \xi~4\left(1-\xi^4\right)^{-\frac{1}{2}}
			~\left\{\Theta\left[\frac{2\mu}{v_{tx}}\sqrt{\frac{2a}{2\mu}}-\xi\right]
			+\Theta\left[\frac{0^{+}}{v_{tx}}\sqrt{\frac{2a}{2\mu}}-\xi\right]\right\}
			\nonumber\\
			&=\Theta\left[\frac{2\mu}{v_{tx}}\sqrt{\frac{a}{\mu}}-1\right]\mathscr{B}\left(1,\frac{1}{4},\frac{1}{2}\right)
			+\Theta\left[1-\frac{2\mu}{v_{tx}}\sqrt{\frac{a}{\mu}}\right]\mathscr{B}\left(\frac{2\mu}{v_{tx}}\sqrt{\frac{a}{\mu}},\frac{1}{4},\frac{1}{2}\right)
			\nonumber\\&
			+\Theta\left[\frac{0^{+}}{v_{tx}}\sqrt{\frac{a}{\mu}}-1\right]\mathscr{B}\left(1,\frac{1}{4},\frac{1}{2}\right)
			+\Theta\left[1-\frac{0^{+}}{v_{tx}}\sqrt{\frac{a}{\mu}}\right]\mathscr{B}\left(\frac{0^{+}}{v_{tx}}\sqrt{\frac{a}{\mu}},\frac{1}{4},\frac{1}{2}\right),
		\end{align}
		and the limit from below
		\begin{align}
			&\lim_{\omega\to 2\mu-0^{+}}\Gamma_{J}\left(\omega,\mu,t_{x}\right)
			\nonumber\\&
			=\frac{1}{2\mathrm{B}\left(\frac{1}{4},\frac{1}{2}\right)}\int_{0}^{+1}d \xi~4\left(1-\xi^4\right)^{-\frac{1}{2}}
			~\left\{\mathrm{sgn}\left(\frac{2\mu-0^{+}}{2}+\mu\right)\Theta\left[\left(\frac{2\mu-0^{+}}{2}+\mu\right)^2-\frac{v_{tx}^2}{2a}(2\mu-0^{+})\xi^{2}\right]
			\right.\nonumber\\&\hspace{5.35cm}\left.
			+\mathrm{sgn}\left(\frac{2\mu-0^{+}}{2}-\mu\right)\Theta\left[\left(\frac{2\mu-0^{+}}{2}-\mu\right)^2-\frac{v_{tx}^2}{2a}(2\mu-0^{+})\xi^2\right]\right\}
			\nonumber\\
			&=\frac{1}{2\mathrm{B}\left(\frac{1}{4},\frac{1}{2}\right)}\int_{0}^{+1}d \xi~4\left(1-\xi^4\right)^{-\frac{1}{2}}
			~\left\{\mathrm{sgn}\left(2\mu-0^{+}\right)\Theta\left[\left(2\mu-0^{+}\right)^2-\frac{v_{tx}^2}{2a}(2\mu)\xi^{2}\right]
			\right.\nonumber\\&\hspace{5.35cm}\left.
			+\mathrm{sgn}\left(-0^{+}\right)\Theta\left[\left(-0^{+}\right)^2-\frac{v_{tx}^2}{2a}(2\mu)\xi^2\right]\right\}
			\nonumber\\
			&=\frac{1}{2\mathrm{B}\left(\frac{1}{4},\frac{1}{2}\right)}\int_{0}^{+1}d \xi~4\left(1-\xi^4\right)^{-\frac{1}{2}}
			~\left\{\Theta\left[\left(2\mu\right)^2-\frac{v_{tx}^2}{2a}(2\mu)\xi^{2}\right]
			-\Theta\left[0^{+}-\frac{v_{tx}^2}{2a}(2\mu)\xi^2\right]\right\}
			\nonumber\\
			&=\frac{1}{2\mathrm{B}\left(\frac{1}{4},\frac{1}{2}\right)}\int_{0}^{+1}d \xi~4\left(1-\xi^4\right)^{-\frac{1}{2}}
			~\left\{\Theta\left[\frac{2\mu}{v_{tx}}\sqrt{\frac{2a}{2\mu}}-\xi\right]
			-\Theta\left[\frac{0^{+}}{v_{tx}}\sqrt{\frac{2a}{2\mu}}-\xi\right]\right\}
			\nonumber\\
			&=\Theta\left[\frac{2\mu}{v_{tx}}\sqrt{\frac{a}{\mu}}-1\right]\mathscr{B}\left(1,\frac{1}{4},\frac{1}{2}\right)
			+\Theta\left[1-\frac{2\mu}{v_{tx}}\sqrt{\frac{a}{\mu}}\right]\mathscr{B}\left(\frac{2\mu}{v_{tx}}\sqrt{\frac{a}{\mu}},\frac{1}{4},\frac{1}{2}\right)
			\nonumber\\&
			-\Theta\left[\frac{0^{+}}{v_{tx}}\sqrt{\frac{a}{\mu}}-1\right]\mathscr{B}\left(1,\frac{1}{4},\frac{1}{2}\right)
			-\Theta\left[1-\frac{0^{+}}{v_{tx}}\sqrt{\frac{a}{\mu}}\right]\mathscr{B}\left(\frac{0^{+}}{v_{tx}}\sqrt{\frac{a}{\mu}},\frac{1}{4},\frac{1}{2}\right).
		\end{align}

		In the untilted phase ($t_{x}=0$), if $\mu\gg 0^{+}$, the limit from above
		\begin{align}
			\lim_{\omega\to 2\mu+0^{+}}\Gamma_{J}\left(\omega,\mu,t_{x}=0\right)
			&=\Theta\left[\frac{2\mu}{0}\sqrt{\frac{a}{\mu}}-1\right]\mathscr{B}\left(1,\frac{1}{4},\frac{1}{2}\right)
			+\Theta\left[1-\frac{2\mu}{0}\sqrt{\frac{a}{\mu}}\right]\mathscr{B}\left(\frac{2\mu}{0}\sqrt{\frac{a}{\mu}},\frac{1}{4},\frac{1}{2}\right)
			\nonumber\\&
			+\Theta\left[\frac{0^{+}}{0}\sqrt{\frac{a}{\mu}}-1\right]\mathscr{B}\left(1,\frac{1}{4},\frac{1}{2}\right)
			+\Theta\left[1-\frac{0^{+}}{0}\sqrt{\frac{a}{\mu}}\right]\mathscr{B}\left(\frac{0^{+}}{0}\sqrt{\frac{a}{\mu}},\frac{1}{4},\frac{1}{2}\right)
			\nonumber\\
			&=\Theta\left[+\infty-1\right]\mathscr{B}\left(1,\frac{1}{4},\frac{1}{2}\right)
			+\Theta\left[1-\infty\right]\mathscr{B}\left(+\infty,\frac{1}{4},\frac{1}{2}\right)
			\nonumber\\&
			+\Theta\left[+\infty-1\right]\mathscr{B}\left(1,\frac{1}{4},\frac{1}{2}\right)
			+\Theta\left[1-\infty\right]\mathscr{B}\left(+\infty,\frac{1}{4},\frac{1}{2}\right)
			\nonumber\\
			&=2\mathscr{B}\left(1,\frac{1}{4},\frac{1}{2}\right),
		\end{align}
		is different from the limit from below
		\begin{align}
			\lim_{\omega\to 2\mu-0^{+}}\Gamma_{J}\left(\omega,\mu,t_{x}=0\right)
			&=\Theta\left[\frac{2\mu}{0}\sqrt{\frac{a}{\mu}}-1\right]\mathscr{B}\left(1,\frac{1}{4},\frac{1}{2}\right)
			+\Theta\left[1-\frac{2\mu}{0}\sqrt{\frac{a}{\mu}}\right]\mathscr{B}\left(\frac{2\mu}{v_{tx}}\sqrt{\frac{a}{\mu}},\frac{1}{4},\frac{1}{2}\right)
			\nonumber\\&
			-\Theta\left[\frac{0^{+}}{0}\sqrt{\frac{a}{\mu}}-1\right]\mathscr{B}\left(1,\frac{1}{4},\frac{1}{2}\right)
			-\Theta\left[1-\frac{0^{+}}{0}\sqrt{\frac{a}{\mu}}\right]\mathscr{B}\left(\frac{0^{+}}{0}\sqrt{\frac{a}{\mu}},\frac{1}{4},\frac{1}{2}\right)
			\nonumber\\
			&=\Theta\left[+\infty-1\right]\mathscr{B}\left(1,\frac{1}{4},\frac{1}{2}\right)
			+\Theta\left[1-\infty\right]\mathscr{B}\left(+\infty,\frac{1}{4},\frac{1}{2}\right)
			\nonumber\\&
			-\Theta\left[+\infty-1\right]\mathscr{B}\left(1,\frac{1}{4},\frac{1}{2}\right)
			-\Theta\left[1-\infty\right]\mathscr{B}\left(+\infty,\frac{1}{4},\frac{1}{2}\right)
			=0.
		\end{align}
		
		In brief, 
		\begin{align}
			\lim_{\omega\to 2\mu+0^{+}}\Gamma_{J}\left(\omega,\mu,t_{x}=0\right)=2\mathscr{B}\left(1,\frac{1}{4},\frac{1}{2}\right)
			&\neq \lim_{\omega\to 2\mu-0^{+}}\Gamma_{J}\left(\omega,\mu,t_{x}=0\right)=0,
		\end{align}
		which indicates that the interband optical transition is allowed when $\omega>2\mu$ but forbidden when $\omega<2\mu$ due to the Pauli exclusion principle and energy conservation.

		In the tilted phase ($t_{x}\gg 0^{+}$), if $\mu\gg 0^{+}$, the limit from above is the same to the limit from below, namely,
		\begin{align}
			\lim_{\omega\to 2\mu+0^{+}}\Gamma_{J}\left(\omega,\mu,t_{x}\right)
			&=\lim_{\omega\to 2\mu-0^{+}}\Gamma_{J}\left(\omega,\mu,t_{x}\right)
			\nonumber\\&
			=\Gamma_{J}\left(2\mu,\mu,t_{x}\right)
			=\Theta\left[\frac{1}{t_{x}}-1\right]\mathscr{B}\left(1,\frac{1}{4},\frac{1}{2}\right)
			+\Theta\left[1-\frac{1}{t_{x}}\right]\mathscr{B}\left(\frac{1}{t_{x}},\frac{1}{4},\frac{1}{2}\right).
		\end{align}
		
		In addition, we derive the expression of $\Gamma_{J}\left(2\mu,\mu,v_{tx}=v_{xc}\right)$ by taking the limit from above
		\begin{align}
			\lim_{t_{x}\to 1+0^{+}}\Gamma_{J}\left(2\mu,\mu,t_{x}\right)
			&=\lim_{t_{x}\to 1+0^{+}}
			\left\{\Theta\left[\frac{1}{t_{x}}-1\right]\mathscr{B}\left(1,\frac{1}{4},\frac{1}{2}\right)
			+\Theta\left[1-\frac{1}{t_{x}}\right]\mathscr{B}\left(\frac{1}{t_{x}},\frac{1}{4},\frac{1}{2}\right)\right\}
			\nonumber\\
			&=\Theta\left[\frac{1}{1+0^{+}}-1\right]\mathscr{B}\left(1,\frac{1}{4},\frac{1}{2}\right)
			+\Theta\left[1-\frac{1}{1+0^{+}}\right]\mathscr{B}\left(\frac{1}{1+0^{+}},\frac{1}{4},\frac{1}{2}\right)
			\nonumber\\&
			=\Theta\left[-0^{+}\right]\mathscr{B}\left(1,\frac{1}{4},\frac{1}{2}\right)
			+\Theta\left[+0^{+}\right]\mathscr{B}\left(1-0^{+},\frac{1}{4},\frac{1}{2}\right)
			\nonumber\\&
			=\Theta\left[+0^{+}\right]\mathscr{B}\left(1-0^{+},\frac{1}{4},\frac{1}{2}\right)
			=\mathscr{B}\left(1,\frac{1}{4},\frac{1}{2}\right),
		\end{align}
		and the limit from below
		\begin{align}
			\lim_{t_{x}\to 1-0^{+}}\Gamma_{J}\left(2\mu,\mu,t_{x}\right)
			&=\lim_{t_{x}\to 1-0^{+}}
			\left\{\Theta\left[\frac{1}{t_{x}}-1\right]\mathscr{B}\left(1,\frac{1}{4},\frac{1}{2}\right)
			+\Theta\left[1-\frac{1}{t_{x}}\right]\mathscr{B}\left(\frac{1}{t_{x}},\frac{1}{4},\frac{1}{2}\right)\right\}
			\nonumber\\
			&=\Theta\left[\frac{1}{1-0^{+}}-1\right]\mathscr{B}\left(1,\frac{1}{4},\frac{1}{2}\right)
			+\Theta\left[1-\frac{1}{1-0^{+}}\right]\mathscr{B}\left(\frac{1}{1-0^{+}},\frac{1}{4},\frac{1}{2}\right)
			\nonumber\\&
			=\Theta\left[+0^{+}\right]\mathscr{B}\left(1,\frac{1}{4},\frac{1}{2}\right)
			+\Theta\left[-0^{+}\right]\mathscr{B}\left(1+0^{+},\frac{1}{4},\frac{1}{2}\right)
			\nonumber\\&
			=\mathscr{B}\left(1,\frac{1}{4},\frac{1}{2}\right),
		\end{align}
		which indicates that the limit from above is the same as the limit from below.
		
		In short,  we have the explicit expressions of $\Gamma_{J} (2\mu+0^{+},\mu,t_{x})$ for all Lifshitz phases as
		\begin{align}
			\Gamma_{J} (2\mu+0^{+},\mu,t_{x}) 
			=
			\begin{cases}
				2\mathscr{B}\left(1,\frac{1}{4},\frac{1}{2}\right),&t_{x}=0,\\\\
				\mathscr{B}\left(1,\frac{1}{4},\frac{1}{2}\right),&0<t_{x}<1,\\\\
				\mathscr{B}\left(1,\frac{1}{4},\frac{1}{2}\right),&t_{x}=1,\\\\
				\mathscr{B}\left(\frac{1}{t_{x}},\frac{1}{4},\frac{1}{2}\right),&t_{x}>1,
			\end{cases}
		\end{align}
		with $\mathscr{B}\left(1,\frac{1}{4},\frac{1}{2}\right)=\frac{1}{2}$.
		
		Consequently, we have
		\begin{align}
			\Gamma_{J}(2\mu+0^{+},\mu,0<t_{x}\le 1)=\frac{\Gamma_{J}(2\mu+0^{+},\mu,t_{x}=0)}{2},
		\end{align}
		which hence leads to 
		\begin{align}
			\mathcal{J}(2\mu+0^{+},\mu,0<t_{x}\le 1)=\frac{\mathcal{J}(2\mu+0^{+},\mu,t_{x}=0)}{2}.
		\end{align}

	\end{widetext}
	
	%%The end of the Appendices%%
\bibliography{LOC.bib}
\end{document}